\documentclass[12pt]{article}%
\usepackage{amsmath,latexsym}
\usepackage{graphicx}
\usepackage{amsmath}
\usepackage{amsfonts}
\usepackage{amssymb}%
\providecommand{\U}[1]{\protect\rule{.1in}{.1in}}
\begin{document}

\title{{\Huge On the energy of Ho\v{r}ava-Lifshitz black holes}}
\author{{\large I. Radinschi \thanks{radinschi@yahoo.com}, F.
Rahaman\thanks{farook\_rahaman@yahoo.com}, and A. Banerjee
\thanks{ayan\_7575@yahoo.co.in}}
\and $^{\ast}${\small Department of Physics, "Gh. Asachi" Technical University,
Iasi, 700050, Romania }
\and $^{\dag}$ {\small Department of Mathematics, Jadavpur University, Kolkata -
700032, India}
\and $^{\ddag}${\small Dept. of Maths, Adamas Institute of Technology, Barasat,
North 24 Parganas - 700126, India }}
\maketitle
\date{}

\begin{abstract}
In this paper we calculate the energy distribution of the Mu-in Park,
Kehagias-Sfetsos (KS) and L\"{u}, Mei and Pope (LMP) black holes in the
Ho\v{r}ava-Lifshitz theory of gravity. These black hole solutions correspond
to the standard Einstein-Hilbert action in the infrared limit. For our
calculations we use the Einstein and M\o ller prescriptions. Various limiting
and particular cases are also discussed.

\end{abstract}

\section{Introduction}

Gravitational energy localization is a very interesting topic of general
relativity theory, which has been tackled by many researchers over the years.
Regardless of the type of approach and of the mathematical methods employed,
there has not been yet developed a generally accepted expression of the
gravitational energy density. This is the problem that needs to be solved and
we use this opportunity to enumerate several definitions able to settle it,
from superenergy tensors [1], quasi local quantities [2], energy-momentum
complexes [3]-[9] and up to teleparallel theory of gravitation (TEGR) [10].
The prescriptions of Einstein [3], Landau and Lifshitz [4], Bergmann-Thomson
[5], Qadir-Sharif [6], Weinberg [7], Papapetrou [8] and M\o ller [9] have been
successfully used by the pseudotensorial theory for the evaluation of
energy-momentum of various gravitational backgrounds. A common characteristic
of the first six definitions is the fact that quasi-Cartesian-coordinates need
to be used to calculate the energy-momentum. As for M\o ller's prescriptions,
studies may be conducted in any system of coordinates.

\qquad As we have noted the use of pseudotensorial definitions to calculate
energy-momentum has led to very good results for several space-times. One
should also bear in mind the results obtained for the $3+1$, $2+1$ and $1+1$
dimensional space-times [11]. In recent years, the similar results provided by
the pseudotensorial method and the teleparallel theory of gravitation have
also been of the highest importance [12].

\qquad As concerns M\o ller's prescription, its use for the calculation of the
energy-momentum of a given gravitational background is supported both by
Cooperstock's important assumption [13] and by Lessner's opinion [14].
Einstein's and M\o ller's prescriptions have had very good results in
gravitational energy localization, which have also been supported by the
numerous interesting works of the last few years [15] and by the quasi-local
theory recently defined by Chang, Nester and Chen [16], which establishes a
direct connection between quasi-local quantities and pseudotensors.

The remainder of our paper is structured as follows: in Section 2 we present
the solutions given by Mu-In Park [17], Kehagias-Sfetsos (KS) [18] and L\"{u},
Mei and Pope (LMP) [19] in the Ho\v{r}ava-Lifshitz theory, and that correspond
to the case of the Einstein-Hilbert action in the IR limit. We also present
the Einstein and M\o ller energy-momentum complexes and calculate the
energy-momentum for the Mu-in Park, (KS) and (LMP) black hole solutions. In
Discussion we point out some limiting and particular cases. For the
calculations we choose the signature ($1,-1,-1,-1$), the geometrized units
($c=1;G=1$) and consider that Greek (Latin) indices take value from $0$ to $3$
and $1$ to $3$.

\section{\bigskip Energy-Momentum of the Mu-In Park, Kehagias-Sfetsos (KS) and
L\"{u}, Mei and Pope\ (LMP) Black Hole Solutions}

We start this section by presenting the three black hole solutions that we use
for our calculations and, after this, the pseudotensorial definitions that we
employed for performing the evaluation of the energy-momentum.

Recently Ho\v{r}ava has proposed a renormalizable gravity theory [20] with
higher spatial derivatives in four dimensions. This theory may be regarded as
a UV complete candidate for general relativity. The theory comes back to
Einstein gravity with a non-vanishing cosmological constant in IR, but it has
improved UV behaviors. The Ho\v{r}ava-Lifshitz theory has been considered very
interesting and after its formulation many researchers found out new black
hole solutions [17, [18, [19] and [21] (and the references there in) and a lot
of work has been done in connection with the Ho\v{r}ava-Lifshitz theory [22].

Consider a static and spherically symmetric solution given by
\begin{equation}
ds^{2}=e^{\nu(r)}dt^{2}-e^{\lambda(r)}dr^{2}-r^{2}(d\theta^{2}+sin^{2}\theta
d\phi^{2}), \tag{1}%
\end{equation}
where the functions$\ \nu(r)$ and $\lambda(r)$ are the metric potentials.

Now by imposing $\lambda_{g}=1$, which reduces to the Einstein-Hilbert action
in the infra-red limit, the following solution of the vacuum field equations
in Ho\v{r}ava gravity [17] is obtained:%

\begin{equation}
e^{\nu(r)}=e^{-\lambda(r)}=1+(w-\Lambda_{W})r^{2}-\sqrt{r[w(w-2\Lambda
_{W})r^{3}+\beta]}. \tag{2}%
\end{equation}
Here $\beta$ is an integration constant and $\lambda_{g}$, $\Lambda_{W}$ and
$w$ are constant parameters. Now the Kehagias-Sfetsos (KS) [18] black hole
solution is obtained by considering $\beta=4wM$ and $\Lambda_{W}=0$
\begin{equation}
e^{\nu(r)}=1+wr^{2}-wr^{2}\sqrt{1+\frac{4M}{wr^{3}}}. \tag{3}%
\end{equation}

By considering $\beta=-\frac{\alpha^{2}}{\Lambda_{W}}$ and $w=0$ the solution
given by Eq. (2) reduces to the L\"{u}, Mei and Pope (LMP) [19] solution,
given by%

\begin{equation}
e^{\nu(r)}=1-\Lambda_{W}r^{2}-\frac{\alpha}{\sqrt{-\Lambda_{W}}}\sqrt{r}.
\tag{4}%
\end{equation}

In the following, we present the Einstein and M\o ller energy-momentum complexes.

Einstein's energy-momentum complex [3] in a four-dimensional space-time is
given by:
\begin{equation}
\theta_{\nu}^{\mu}=\frac{1}{16\pi}h_{\nu,\,\lambda}^{\mu\lambda}. \tag{5}%
\end{equation}
The Einstein superpotential $h_{\nu}^{\mu\lambda}$ has the expression:%
\begin{equation}
h_{\nu}^{\mu\lambda}=\frac{1}{\sqrt{-g}}g_{\nu\sigma}[-g(g^{\mu\sigma
}g^{\lambda\kappa}-g^{\lambda\sigma}g^{\mu\kappa})]_{,\kappa} \tag{6}%
\end{equation}
and presents the antisymmetry property
\begin{equation}
h_{\nu}^{\mu\lambda}=-h_{\nu}^{\lambda\mu}. \tag{7}%
\end{equation}
$\theta_{0}^{0}$ and $\theta_{i}^{0}$ are the energy and momentum density
components, respectively. The Einstein energy-momentum complex satisfies the
local conservation law
\begin{equation}
\theta_{\nu,\,\mu}^{\mu}=0. \tag{8}%
\end{equation}
The energy and momentum in Einstein's prescription are given by
\begin{equation}
P_{\mu}=\int\int\int\theta_{\mu}^{0}\,dx^{1}dx^{2}dx^{3} \tag{9}%
\end{equation}
and the energy of a physical system in a four-dimensional background is
\begin{equation}
E=\int\int\int\theta_{0}^{0}dx^{1}dx^{2}dx^{3}. \tag{10}%
\end{equation}
In eq. (9) $P_{i}$, $i=1,2,3$, are the momentum components. In the case of the
Einstein definition the calculations are restricted to quasi-Cartesian
coordinates. For performing the calculations we have to transform the metric
given by (1) in Schwarzschild Cartesian coordinates, as given by%

\begin{equation}
ds^{2}=Ndt^{2}-(dx^{2}+dy^{2}+dz^{2})-\frac{N^{-1}-1}{r^{2}}(xdx+ydy+zdz)^{2},
\tag{11}%
\end{equation}

where $N$ is employed by $e^{\nu(r)}=e^{-\lambda(r)}$ corresponding to the
Mu-In Park, (KS) and (LMP) black hole solutions, respectively. We have
$N=1+(w-\Lambda_{W})r^{2}-\sqrt{r[w(w-2\Lambda_{W})r^{3}+\beta]}$ for the
Mu-In Park black hole solution, $N=$ $1+wr^{2}-wr^{2}\sqrt{1+\frac{4M}{wr^{3}%
}}$ for the Kehagias-Sfetsos (KS) gravitational background and $N=-\Lambda
_{W}r^{2}-\frac{\alpha}{\sqrt{-\Lambda_{W}}}\sqrt{r}$ for the L\"{u}, Mei and
Pope (LMP) black hole solution.

For the Mu-In Park gravitational background described by (2), with
$N=1+(w-\Lambda_{W})r^{2}-\sqrt{r[w(w-2\Lambda_{W})r^{3}+\beta]}$ and using
(10) and (11) we obtain%

\begin{equation}
E(r)=\frac{r}{2}\left[  -(w-\Lambda_{W})r^{2}+\sqrt{r[w(w-2\Lambda_{W}%
)r^{3}+\beta]}\right]  . \tag{12}%
\end{equation}

In the case of the (KS) black hole solution described by (3) and with $N=$
$1+wr^{2}-wr^{2}\sqrt{1+\frac{4M}{wr^{3}}}$ we compute the energy distribution
with the aid of (10) and (11) and we get%

\begin{equation}
E(r)=\frac{r}{2}\left[  wr^{2}\left(  -1+\sqrt{1+\frac{4M}{wr^{3}}}\right)
\right]  . \tag{13}%
\end{equation}

\begin{figure}[ptb]
\begin{center}
\vspace{0.5cm} \includegraphics[width=0.5\textwidth]{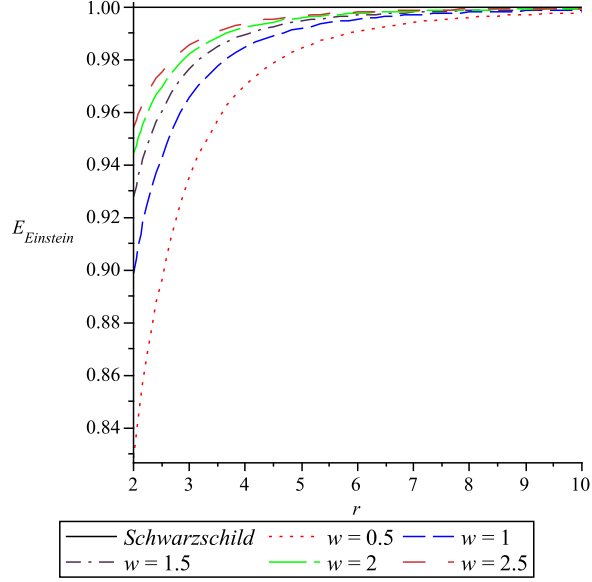}
\end{center}
\caption{The plot for the Einstein energy E vs. r, in the case of the (KS)
black hole solution, for various values of w and with M =1.}%
\label{fig1}%
\end{figure}

For the (LMP) black hole solution using $N=-\Lambda_{W}r^{2}-\frac{\alpha
}{\sqrt{-\Lambda_{W}}}\sqrt{r}$ (10) and (11) the energy distribution is given by%

\begin{equation}
E(r)=\frac{r}{2}\left[  \Lambda_{W}r^{2}+\frac{\alpha}{\sqrt{-}\Lambda_{W}%
}\sqrt{r}\right]  . \tag{14}%
\end{equation}

In the case of these black hole solutions all the momenta are found to be zero.

The M{\o }ller energy-momentum complex is defined by%

\begin{equation}
\mathcal{J}_{\nu}^{\mu}=\frac{1}{8\pi}M_{\nu\,\,,\,\lambda}^{\mu\lambda},
\tag{15}%
\end{equation}
with the M{\o }ller superpotential $M_{\nu}^{\mu\lambda}$ given by
\begin{equation}
M_{\nu}^{\mu\lambda}=\sqrt{-g}\left(  \frac{\partial g_{\nu\sigma}}{\partial
x^{\kappa}}-\frac{\partial g_{\nu\kappa}}{\partial x^{\sigma}}\right)
g^{\mu\kappa}g^{\lambda\sigma}, \tag{16}%
\end{equation}
which presents the antisymmetric property
\begin{equation}
M_{\nu}^{\mu\lambda}=-M_{\nu}^{\lambda\mu}. \tag{17}%
\end{equation}

M{\o }ller's energy-momentum complex satisfies the local conservation law
\begin{equation}
\frac{\partial\mathcal{J}_{\nu}^{\mu}}{\partial x^{\mu}}=0, \tag{18}%
\end{equation}
where $\mathcal{J}_{0}^{0}$ is the energy density and $\mathcal{J}_{i}^{0}$
are the momentum density components.

In the M{\o }ller prescription the energy and momentum are given by
\begin{equation}
P_{\mu}=\int\int\int\mathcal{J}_{\mu}^{0}dx^{1}dx^{2}dx^{3}. \tag{19}%
\end{equation}
The energy distribution is
\begin{equation}
E=\int\int\int\mathcal{J}_{0}^{0}dx^{1}dx^{2}dx^{3}. \tag{20}%
\end{equation}

Using the Gauss theorem and evaluating the integral over the surface of a
sphere of radius $r$ the expression for energy is given by%

\begin{equation}
E=\frac{1}{8\pi}%
{\displaystyle\oint\limits_{r}}
M_{0}^{01}\sin\theta d\theta d\varphi. \tag{21}%
\end{equation}

The M\o ller definition allows to make the calculations in any coordinate
system, and for our purpose we use the metrics given by (2), (3) and (4). For
the Mu-in Park black hole solution the non-zero components of the M\o ller
energy-momentum complex are%

\begin{equation}
M_{2}^{21}=-2r\sin\theta(-1-r^{2}w+r^{2}\Lambda_{W}+\sqrt{r[w(w-2\Lambda
_{W})r^{3}+\beta]}), \tag{22}%
\end{equation}

\begin{equation}
M_{3}^{31}=-2r\sin\theta(-1-r^{2}w+r^{2}\Lambda_{W}+\sqrt{r[w(w-2\Lambda
_{W})r^{3}+\beta]}), \tag{23}%
\end{equation}

\begin{equation}
M_{3}^{32}=2\cos\theta, \tag{24}%
\end{equation}

\begin{equation}
M_{0}^{01}=-\frac{1}{2}\frac{r^{2}\sin\theta\lbrack4r\sqrt{r[w(w-2\Lambda
_{W})r^{3}+\beta]}(\Lambda_{W}-w)+4w(w-2\Lambda_{W})r^{3}+\beta]}%
{\sqrt{r[w(w-2\Lambda_{W})r^{3}+\beta]}}. \tag{25}%
\end{equation}

Using (21) and (25) we obtain the expression for the energy distribution of
the Mu-In Park black hole solution that is given by%

\begin{equation}
E(r)=\frac{r^{2}}{2}\left[  2(w-\Lambda_{W})r-\frac{1}{2}\frac{4w(w-2\Lambda
_{W})r^{3}+\beta}{\sqrt{r[w(w-2\Lambda_{W})r^{3}+\beta]}}\right]  . \tag{26}%
\end{equation}

In the case of the (KS) gravitational background the non-zero M\o ller
superpotentials are%

\begin{equation}
M_{2}^{21}=-2r\sin\theta\left[  -1-wr^{2}\left(  1-\sqrt{1+\frac{4M}{wr^{3}}%
}\right)  \right]  , \tag{27}%
\end{equation}

\begin{equation}
M_{3}^{31}=-2r\sin\theta\left[  -1-wr^{2}\left(  1-\sqrt{1+\frac{4M}{wr^{3}}%
}\right)  \right]  , \tag{28}%
\end{equation}

\begin{equation}
M_{3}^{32}=2\cos\theta, \tag{29}%
\end{equation}

\begin{equation}
M_{0}^{01}=2\sin\theta\frac{\left[  wr^{3}\left(  -1+\sqrt{1+\frac{4M}{wr^{3}%
}}\right)  -M\right]  }{\sqrt{1+\frac{4M}{wr^{3}}}}. \tag{30}%
\end{equation}

For the (KS) black hole solution given by (3) we insert (30) into (21) and we get%

\begin{equation}
E(r)=\frac{\left[  wr^{3}\left(  -1+\sqrt{1+\frac{4M}{wr^{3}}}\right)
-M\right]  }{\sqrt{1+\frac{4M}{wr^{3}}}}. \tag{31}%
\end{equation}

\begin{figure}[ptb]
\begin{center}
\vspace{0.5cm} \includegraphics[width=0.5\textwidth]{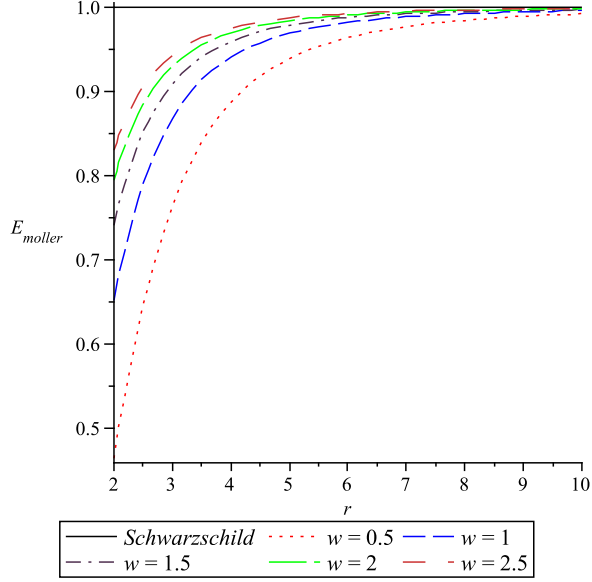}
\end{center}
\caption{The plot for the M\o ller energy vs. r, in the case of the (KS) black
hole solution, for various values of w and with M =1.}%
\label{fig1}%
\end{figure}

The non-zero components of the M\o ller energy-momentum complex in the case of
the (LMP) black hole solution are%

\begin{equation}
M_{2}^{21}=-2\frac{r\sin\theta(-\sqrt{-\Lambda_{W}}+\Lambda_{W}r^{2}%
\sqrt{-\Lambda_{W}}+\alpha\sqrt{r})}{\sqrt{-\Lambda_{W}}}, \tag{32}%
\end{equation}

\begin{equation}
M_{3}^{31}=-2\frac{r\sin\theta(-\sqrt{-\Lambda_{W}}+\Lambda_{W}r^{2}%
\sqrt{-\Lambda_{W}}+\alpha\sqrt{r})}{\sqrt{-\Lambda_{W}}}, \tag{33}%
\end{equation}

\begin{equation}
M_{3}^{32}=2\cos\theta, \tag{34}%
\end{equation}

\begin{equation}
M_{0}^{01}=-\frac{1}{2}\frac{r^{2}\sin\theta(4\Lambda_{W}r^{3/2}\sqrt
{-\Lambda_{W}}+\alpha)}{\sqrt{-\Lambda_{W}}\sqrt{r}}. \tag{35}%
\end{equation}

In the case of the (LMP) black hole solution described by (4) using (35) and
(21) we obtain the energy distribution%

\begin{equation}
E(r)=\frac{r^{2}}{2}\left[  -2\Lambda_{W}r-\frac{1}{2}\frac{\alpha}{\sqrt
{r}\sqrt{-\Lambda_{W}}}\right]  . \tag{36}%
\end{equation}

For the Mu-In Park, Kehagias-Sfetsos (KS) and L\"{u}, Mei and Pope (LMP) black
hole solutions all the momenta are found to be zero.

\section{Discussion}

In this paper we calculate the energy distribution of Mu-In Park,
Kehagias-Sfetsos (KS) and L\"{u}, Mei and Pope (LMP) black holes in the
Einstein and M\o ller prescriptions. For the gravitational background
described by the Mu-in Park metric we found that the energy distribution
depends on the $w$, $\Lambda_{W}$ parameters and $\beta$ both in the Einstein
and M\o ller prescriptions. For the space-time given by the Kehagias-Sfetsos
(KS) metric the energy depends on $w$ and the mass $M$. In the case of the
L\"{u}, Mei and Pope (LMP) black hole solution the energy distribution
presents a dependence in function of the $\Lambda_{W}$ parameter and $\alpha$.
In both prescriptions for these gravitational backgrounds all the momenta are zero.

We present some particular and limiting cases.

For the Mu-In Park black hole solution a particular case is obtained for
$r>>[\beta/w(w-2\Lambda_{W})]^{1/3}$ when $N$ has a new expression
$N=1+\frac{\Lambda_{W}^{2}}{2w}r^{2}-\frac{\beta}{2\sqrt{w(w-2\Lambda_{W})}%
}\frac{1}{r}+O(r^{-4})$. This condition combined with $\Lambda_{W}=0$ and
$\beta=4wM$ independently of $w$ leads to the usual behaviour of a
Schwarzschild black hole solution. Using (11) with the new expression for $N$
and (10) the energy distribution in the Einstein prescription is
$E=M-O(r^{-3})$. The M\o ller definition gives for the energy distribution the
expression $E=M-O(r^{-3})$. For large $r$ both prescriptions yield the same
result $E=M$, which is also equal to the ADM mass. These are expected results
in the context of general relativity (at large distances the standard general
relativity is recovered).

In the case of the (KS) gravitational background for $r>>(M/\omega)^{1/3}$ the
usual behavior of a Schwarzschild black hole is obtained. For large $r$ the
Einstein and M\o ller definitions yield for the energy distribution the
expression $E=M$.

In the following, we study the case $r\rightarrow\infty$ for the Mu-In Park,
Kehagias-Sfetsos (KS) and L\"{u}, Mei and Pope (LMP) gravitational
backgrounds. The results for the Einstein prescription are presented in Table
1.%
\[%
\begin{tabular}
[c]{lll}%
The Einstein prescription~~~ & ~~~~~~~~~~~~~~~Energy ~~~~~ & ~~~~$r$\\
&  & \\
Mu-In Park & $signum(-\frac{1}{2}w+\frac{1}{2}\Lambda_{W}+\frac{1}{2}%
\sqrt{w^{2}-2w\Lambda_{W}})~\infty$~~~~ & $r\rightarrow\infty$\\
&  & \\
Kehagias-Sfetsos (KS)~~~~ & ~~~~~~~~~ $M$ & $r\rightarrow\infty$\\
&  & \\
L\"{u}, Mei and Pope (LMP) ~~~ & ~~~ $signum(\Lambda_{W})~\infty$~~~~~ &
$r\rightarrow\infty$%
\end{tabular}
\ \ \
\]

\[
\text{Table 1}%
\]

The limiting cases for large $r$ in the case of the M\o ller prescription are
given in Table 2.%

\[%
\begin{tabular}
[c]{lll}%
The M\o ller prescription~~~~~~~~~~~~~ & ~~~~~~~~~Energy ~~~ & $r$\\
&  & \\
Mu-In Park & $-signum(\frac{-\sqrt{w^{2}-2w\Lambda_{W}}(w-\Lambda
_{W})-2w\Lambda_{W}+w^{2})}{\sqrt{w^{2}-2w\Lambda_{W}}})~\infty$ &
$r\rightarrow\infty$\\
&  & \\
Kehagias-Sfetsos (KS) & ~~~~~~~~~~~~~$M$ & $r\rightarrow\infty$\\
&  & \\
L\"{u}, Mei and Pope (LMP) & $-signum(\Lambda_{W})~\infty$ & $r\rightarrow
\infty$%
\end{tabular}
\ \ \
\]

\[
\text{Table 2}%
\]

\begin{figure}[ptb]
\begin{center}
\vspace{0.5cm} \includegraphics[width=0.5\textwidth]{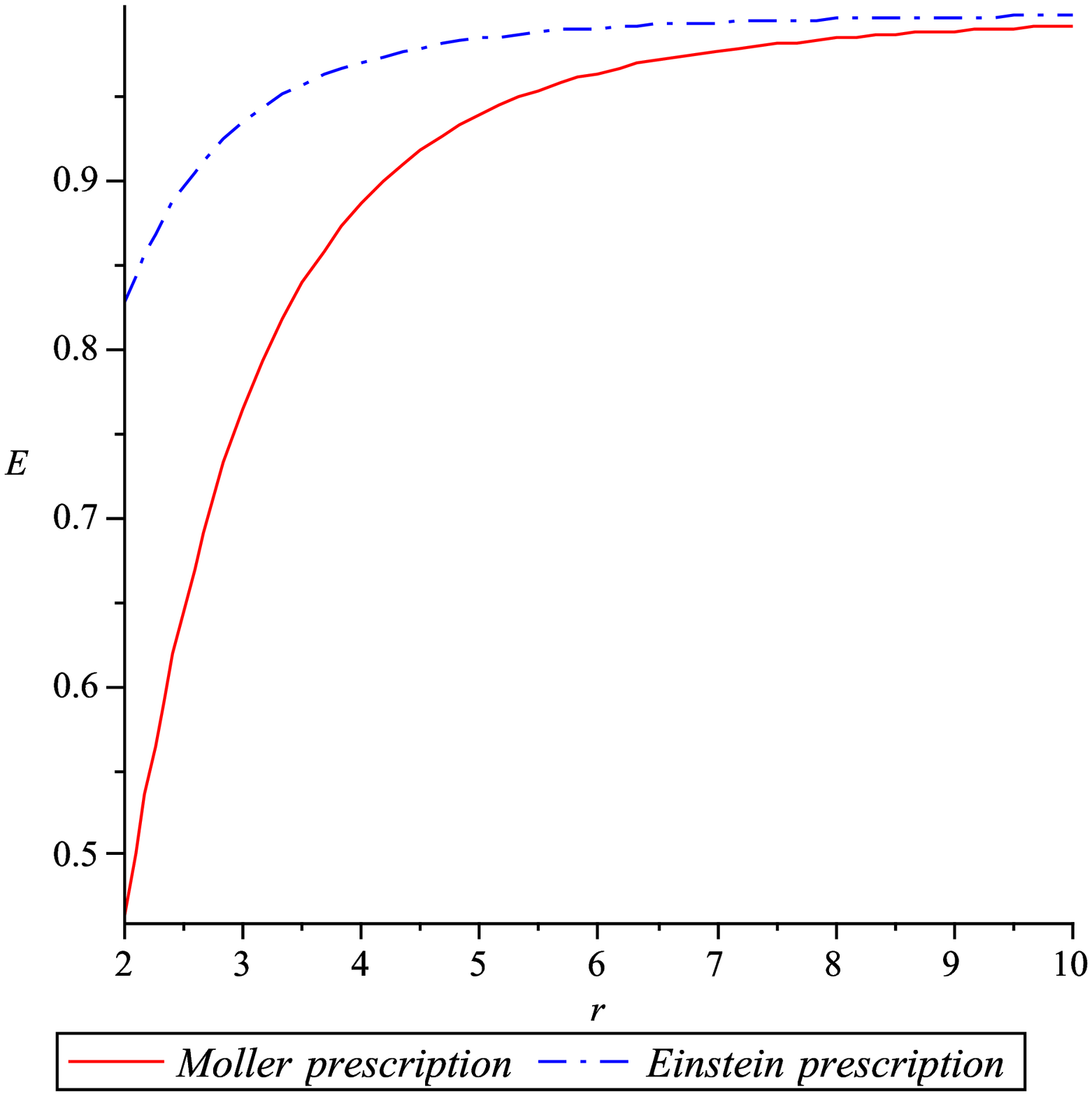}
\end{center}
\caption{ Energy distributions in the Einstein and M\o ller's prescription vs.
r for small values of w like w =.5 and with M =1.}%
\label{fig1}%
\end{figure}

\begin{figure}[ptb]
\begin{center}
\vspace{0.5cm} \includegraphics[width=0.5\textwidth]{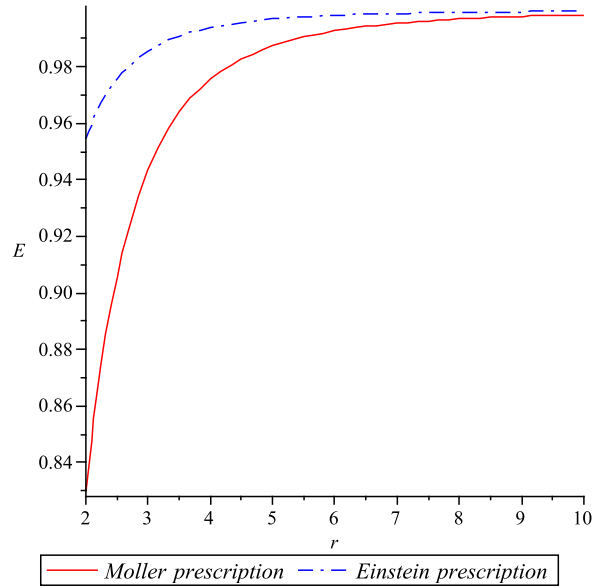}
\end{center}
\caption{Energy distributions in the Einstein and M\o ller's prescription vs.
r for large values of w like w =2.5 and with M =1.}%
\label{fig1}%
\end{figure}

In Fig. 3 and Fig. 4 are plotted the energy distributions in the Einstein and
M\o ller's prescription for the (KS) black hole solution in two cases. In Fig.
3 we have the plot of the energy distributions vs. $r$ for small values of $w$
like $w=.5$ and with $M=1$. In Fig. 4 we present the plot of the energy
distributions vs.$r$ for large value of $w$ like $w=2.5$ and with $M=1$.

An interesting conclusion is that for small values of $r$, the amount of
energy can be much smaller for a Kehagias-Sfetsos (KS) black hole obtained by
using both Einstein and M\o ller's prescriptions than for a Schwarzschild
black hole. However, for large $r$ both the Einstein and M\o ller
prescriptions give the same expression for energy distribution $E=M$. In the
case of general relativity this represents the ADM mass. This result also
confirms the fact that at large distances the Ho\v{r}ava-Lifshitz theory leads
to the Einstein gravity, in the context of a particular coupling $\lambda
_{g}=1$ and both the Einstein and M\o ller prescriptions yield the same
expression for energy, as is expected in this case. One can also note that in
the case of small values of $r$ the energy of Kehagias-Sfetsos (KS) black hole
solution obtained by using Einstein's prescription is greater than the energy
distribution obtained by using M\o ller's prescription. This difference is
decreasing with the increasing of the values of $w$.

One can note that for $r\rightarrow0$ the energy $E\rightarrow0$. But in
realistic situation, one should consider the value of $r>r_{h}$ ( horizon).

The black hole in Ho\v{r}ava-Lifshitz\ gravity in Mu-In Park gravitational
background is more general than KS and LPM solutions. The solution contains a
number of parameters and for specific choices of the parameters, one could
come back to KS and LPM solutions. Also, one can see that for large $w$, the
KS case comes back to the Schwarzschild case (Einstein gravity). Also, we have
shown a comparison between the two gravitational theories, based mostly on the plots.

The results that we obtained for the energy in the case of the
Ho\v{r}ava-Lifshitz black hole space-times demonstrate that the
pseudotensorial prescriptions are useful concepts for the evaluation of the
energy and momentum\textbf{.}

\end{document}